\documentclass[11pt,english]{article}
\usepackage[T1]{fontenc}
\usepackage[latin9]{inputenc}
\usepackage{geometry}
\usepackage{color}
\usepackage{amsmath}
\usepackage{amssymb}
\usepackage{setspace}
\usepackage[authoryear]{natbib}
\makeatletter

\providecommand{\tabularnewline}{\\}

\@ifundefined{date}{}{\date{}}
\date{}
\usepackage{graphicx}

\newtheorem{assumption}{Assumption}
\newtheorem{theorem}{Theorem}

\newtheorem{lemma}{Lemma}

\newtheorem{remark}{Remark}

\newcommand*{\indep}{%
  \mathbin{%
    \mathpalette{\@indep}{}%
  }%
}
\newcommand*{\nindep}{%
  \mathbin{
    \mathpalette{\@indep}{\not}
  }%
}
\newcommand*{\@indep}[2]{%
  \sbox0{$#1\perp\m@th$}
  \sbox2{$#1=$}
  \sbox4{$#1\vcenter{}$}
  \rlap{\copy0}
  \dimen@=\dimexpr\ht2-\ht4-.2pt\relax
  \kern\dimen@
  {#2}%
  \kern\dimen@
  \copy0 
} 

\newcommand{\pr}{P} 

\newcommand{\var}{\mathrm{var}}

\newcommand{\de}{\mathrm{d}}
\newcommand{\T}{\mathrm{\scriptscriptstyle T}}
\newcommand{\logit}{\text{logit}}

\newcommand{\ipw}{\mathrm{ipw}}

\newcommand{\gam}{\mathrm{GAM}}

\newcommand{\nni}{\mathrm{nni}}
\newcommand{\knn}{\mathrm{knn}}

\newcommand{\rc}{\mathrm{RC}}
\newcommand{\HT}{\mathrm{HT}}

\newcommand{\dr}{\mathrm{dr}}

\newcommand{\N}{\mathcal{N}}
\newcommand{\F}{\mathcal{F}}

\usepackage{babel}

\usepackage{babel}

\usepackage{babel}

\makeatother

\usepackage{babel}
\begin{document}

\title{\textbf{\huge{}Integration of survey data and big observational data
for finite population inference using mass imputation }}

\author{Shu Yang\thanks{Department of Statistics, North Carolina State University, North Carolina
27695, U.S.A. Email: syang24@ncsu.edu}$\ $ and Jae Kwang Kim\thanks{Department of Statistics, Iowa State University, Iowa 50011, U.S.A.}}
\maketitle
\begin{abstract}
Multiple data sources are becoming increasingly available for statistical
analyses in the era of big data. As an important example in finite-population
inference, we consider an imputation approach to combining a probability
sample with big observational data. Unlike the usual imputation for
missing data analysis, we create imputed values for the whole elements
in the probability sample. Such mass imputation is attractive in the
context of survey data integration \citep{kim2012combining}. We extend
mass imputation as a tool for data integration of survey data and
big non-survey data. The mass imputation methods and their statistical
properties are presented. The matching estimator of \citet{rivers2007sampling}
is also covered as a special case. Variance estimation with mass-imputed
data is discussed. The simulation results demonstrate the proposed
estimators outperform existing competitors in terms of robustness
and efficiency. 

\bigskip{}
 \textit{Keywords}: Calibration weighting; Data fusion; Generalized
additive model; Nearest neighbor imputation; Post stratification;
Statistical matching.
\end{abstract}

\section{Introduction}

In finite population inference, probability sampling is the gold standard
for obtaining a representative sample from the target population.
Because the selection probability is known, the subsequent inference
from a probability sample is often design-based and respect the way
in which the data were collected. However, large-scale survey programs
continually face heightened demands coupled with reduced resources.
Demands include requests for estimates for domains with small sample
sizes and desires for more timely estimates. Simultaneously, program
budget cuts force reductions in sample sizes, and decreasing response
rates make nonresponse bias an important concern. \citet{baker2013summary}
and \citet{keiding2016perils} address the current challenges in using
probability samples for finite population inferences.

To meet the new challenges, statistical offices face the increasing
pressure to utilize convenient but often uncontrolled big data sources,
such as web survey panels and satellite information. While such data
sources provide timely data for a large number of variables and population
elements, they often fail to represent the target population of interest
because of inherent selection biases. 

To address new objectives and utilize modern data sources in statistically
defensible ways, it is important to develop statistical tools for
data integration for combining a probability sample with big observational
data. To achieve this goal, one can apply mass imputation, where the
imputed values are created for the whole elements in the probability
sample. In the usual imputation for missing data analysis, the respondents
in the sample provide a training dataset for developing an imputation
model. In the proposed mass imputation, an independent big data sample
is used as a training dataset, and mass imputation is applied to the
probability sample. While the mass imputation idea for incorporating
information from big data is very natural, the literature on mass
imputation itself is very sparse. \citet{breidt1996two} discuss mass
imputation for two-phase sampling. \citet{kim2012combining} develop
a rigorous theory for mass imputation using two independent probability
samples. \citet{chipperfield2012combining} discuss composite estimation
when one of the surveys is mass imputed. \citet{rivers2007sampling}
proposes a mass imputation approach using nearest neighbor imputation
but the theory is not fully developed. Recently, Kim and Wang (2018),
a technique report available by request from the authors, develop
a theory for mass imputation for big data using a parametric model
approach. However, the parametric model assumptions do not necessarily
hold in practice. In order for mass imputation to be more useful and
practical, the assumptions should be as weak as possible.

In this paper, we first develop a formal framework for mass imputation
incorporating information from big data into a probability sample
and present rigorous asymptotic results. Unlike Kim and Wang (2018),
we do not make strong parametric model assumptions for mass imputation.
Thus, the proposed method is appealing to survey practitioners. Our
framework covers the nearest neighbor imputation estimator of \citet{rivers2007sampling}.
In $\mathsection$ \ref{sec:tools}, we investigate two strategies
for improving the nearest neighbor imputation estimator, one using
$k$ nearest neighbor imputation and the other using generalized additive
models. Secondly, using a novel calibration weighting idea, we propose
an efficient mass imputation estimator and develop its asymptotic
results. The efficiency gain is justified under a purely design-based
framework and no model assumptions are used. The proposed methods
are evaluated through extensive simulation studies based on artificial
data and real-life data from U.S. Census Bureau\textquoteright s Monthly
Retail Trade Survey. 

\section{Basic Setup\label{sec:Basic-Setup}}

\subsection{Notation: two data sources}

Let $\mathcal{F}_{N}=\{(X_{i},Y_{i}):i\in U\}$ with $U=\{1,\ldots,N\}$
denote a finite population, where $X_{i}=(X_{i}^{1},\ldots,X_{i}^{p})$
is a $p$-dimensional vector of covariates, and $Y_{i}$ is the study
variable. We assume that $\F_{N}$ is a random sample from a superpopulation
model $\zeta$, and $N$ is known. Our objective is to estimate the
general finite population parameter $\mu_{g}=N^{-1}\sum_{i=1}^{N}g(Y_{i})$
for some known $g(\cdot)$. For example, if $g(Y)=Y$, $\mu_{g}=N^{-1}\sum_{i=1}^{N}Y_{i}$
is the population mean of $Y$. If $g(Y)=I(Y<c)$ for some constant
$c$, $\mu_{g}=N^{-1}\sum_{i=1}^{N}I(Y_{i}<c)$ is the population
proportion of $Y$ less than $c$. 

Suppose that there are two data sources, one from a probability sample,
referred to as Sample A, and the other from a big data source, referred
to as Sample B. Table \ref{tab:Two-data-sources} illustrates the
observed data structure. Sample A contains observations $\mathcal{O}_{A}=\{(d_{i}=\pi_{i}^{-1},X_{i}):i\in A\}$
with sample size $n=|A|,$ where $\pi_{i}=\pr(i\in A)$ is known throughout
Sample A, and Sample B contains observations $\mathcal{O}_{B}=\{(X_{i},Y_{i}):i\in B\}$
with sample size $N_{B}=|B|$. Although the big data source has a
large sample size, the sampling mechanism is often unknown, and we
cannot compute the first-order inclusion probability for Horvitz-Thompson
estimation. The naive estimators without adjusting for the sampling
process are subject to selection biases. On the other hand, although
the probability sample with sampling weights represents the finite
population, it does not observe the study variable.

\begin{table}
\begin{centering}
{\scriptsize{}{}\caption{\label{tab:Two-data-sources}Two data sources. ``$\protect\surd$''
and ``?'' indicate observed and unobserved data, respectively. }
}\centering{}%
\begin{tabular}{ccccc}
 &  &  & \multicolumn{1}{c}{} & \tabularnewline
\hline 
 &  & Sample weight & Covariate & Study Variable\tabularnewline
 &  & $d=\pi^{-1}$ & $X$ & $Y$\tabularnewline
\hline 
Probability Sample  & 1  & $\surd$  & $\surd$  & ? \tabularnewline
$\mathcal{O}_{A}$ & $\vdots$  & $\vdots$  & $\vdots$  & $\vdots$ \tabularnewline
 & $n$  & $\surd$  & $\surd$  & ? \tabularnewline
\hline 
Big Data Sample  & 1  & ?  & $\surd$  & $\surd$ \tabularnewline
$\mathcal{O}_{B}$ & $\vdots$  & $\vdots$  & $\vdots$  & $\vdots$ \tabularnewline
 & $N_{B}$  & ?  & $\surd$  & $\surd$ \tabularnewline
\hline 
\end{tabular}
\par\end{centering}
Sample A is a probability sample, and Sample B is a big data but may
have selection biases.
\end{table}

\subsection{Assumptions}

Let $f(Y\mid X)$ be the conditional distribution of $Y$ given $X$
in the superpopulation model $\zeta$. We define $\delta_{B}$ to
be the indicator of selection to Sample B. We first make the following
assumption. 

\begin{assumption}[Ignorability]\label{asmp:MAR}

Conditional on $X$, the distribution of $Y$ in Sample B follows
the superpopulation model; i.e., $f(Y\mid X;\delta_{B}=1)=f(Y\mid X)$. 

\end{assumption}

Assumption \ref{asmp:MAR} states the ignorability of the selection
mechanism to Sample B conditional upon the covariates. This assumption
is also a missingness at random assumption \citep{rubin1976inference}.

Now, let $f(X)$ and $f(X\mid\delta_{B}=1)$ be the density function
of $X$ in the finite population and Sample B, respectively. We also
require the following assumption. 

\begin{assumption}[Common support]\label{asmp:overlap}

The vector of covariates $X$ has a compact and convex support, with
its density bounded and bounded away from zero. There exist constants
$C_{l}$ and $C_{u}$ such that $C_{l}\leq f(X)/f(X\mid\delta_{B}=1)\leq C_{u}$
almost surely.

\end{assumption}

Assumption \ref{asmp:overlap} implies that the support of $X$ in
Sample B is the same as that in the finite population. This assumption
can also be formulated as a positivity assumption that $\pr(\delta_{B}=1\mid X)>0$
for all $X$. This is necessary, because if the probability of selection
into Sample B given some $X$ is zero, then Sample B cannot provide
adequate $Y$ information for the units in this region without extrapolation. 

\section{Methodology\label{sec:Methodology}}

\subsection{Nearest neighbor imputation}

For estimation, if $Y_{i}$ were observed throughout Sample A, the
Horvitz\textendash Thompson estimator $\hat{\mu}_{g,\HT}=N^{-1}\sum_{i\in A}\pi_{i}^{-1}g(Y_{i})$
can be used. Our primary focus will be on the imputation estimator
of $\mu_{g}$, given by $\hat{\mu}_{g,I}=N^{-1}\sum_{i\in A}\pi_{i}^{-1}g(Y_{i}^{*}),$
where $Y_{i}^{*}$ is an imputed value for $Y_{i}$. Creating imputed
values for the whole data is called mass imputation \citep{chipperfield2012combining,kim2012combining}.

To find suitable imputed values, we consider nearest neighbor imputation;
that is, find the closest matching unit from Sample B based on the
$X$ values and use the corresponding $Y$ value from this unit as
the imputed value. This approach has been called statistical matching
by \citet{rivers2007sampling}. To investigate the theoretical properties,
we first consider matching with replacement with single imputation;
the discussion on $k$ nearest neighbor imputation is presented in
$\mathsection$ \ref{sec:tools}.

The nearest neighbor approach to mass imputation can be described
in the following steps: 
\begin{description}
\item [{Step$\ $1.}] For each unit $i\in A$, find the nearest neighbor
from Sample B with the minimum distance between $X_{j}$ and $X_{i}$.
Let $i(1)$ be the index of its nearest neighbor, which satisfies
$d(X_{i(1)},X_{i})\le d(X_{j},X_{i}),$ for $j\in B$, where $d(X_{i},X_{j})$
is a distance function between $X_{i}$ and $X_{j}$. Without loss
of generality, we use the Euclidean distance, $d(X_{i},X_{j})=||X_{i}-X_{j}||$,
where $||X||=(X^{\T}X)^{1/2}$, to determine neighbors; our theoretical
development applies to other distances \citep{abadie2006large}. 
\item [{Step$\ $2.}] The nearest neighbor imputation estimator of $\mu_{g}$
is
\begin{equation}
\hat{\mu}_{g,\nni}=\frac{1}{N}\sum_{i\in A}\pi_{i}^{-1}g(Y_{i(1)}).\label{eq:nni}
\end{equation}
\end{description}
The matching estimator is attractive in practice because it does not
require parametric model assumptions. Secondly, it does not require
Sample A and Sample B to have common units, but requires only Assumption
\ref{asmp:overlap}. Assumption \ref{asmp:overlap} ensures that for
any $X_{i}$ in Sample A, we can find a value $X_{i(1)}$ in Sample
B that is arbitrarily close to $X_{i}$ as $N_{B}\rightarrow\infty$.
Then, by Assumption \ref{asmp:MAR}, $g(Y_{i(1)})$ has the same distribution
of $g(Y_{i})$, given $X_{i}$. Moreover, for the same imputed dataset,
one can estimate different parameters by choosing reasonable $g(\cdot)$.
The main weakness of nearest neighbor imputation is that it is subject
to the curse of dimensionality when $X$ is a vector, but such weakness
is not applicable when the size of the matching donor pool is huge
as in our big data setup.

\subsection{Asymptotic results}

To study the asymptotic properties of $\hat{\mu}_{g,\nni}$, we impose
the following regularity conditions on the functional continuity and
finite moments (e.g., \citealp{mack1981local}) and the sampling design
for Sample A (\citealp{fuller2009sampling}, \textcolor{black}{Ch.
1}).

\begin{assumption}\label{asmp:regularity} (i) $f(X)$ and $\mu_{g}(X)=E\{g(Y)\mid X\}$
are continuously differentiable for any continuous and bounded $g(Y)$,
and (ii) $E\{g(Y)^{\beta}\mid X\}$ is bounded for $\beta=0,1,2$.

\end{assumption} 

\begin{assumption}\label{asmp:sampling} (i) There exist positive
constants $C_{1}$ and $C_{2}$ such that $C_{1}\le Nn^{-1}\pi_{i}\le C_{2},$
for $i=1,\ldots,N$; (ii) the sampling fraction for Sample A is negligible,
$nN^{-1}=o(1)$; and (iii) the sequence of the Horvitz-Thompson  estimators
$\hat{\mu}_{g,\HT}$ satisfies $\var_{p}(\hat{\mu}_{g,\HT})=O(n^{-1})$
and $\{\var_{p}(\hat{\mu}_{g,\HT})\}^{-1/2}(\hat{\mu}_{g,\HT}-\mu_{g})\mid\mathcal{F}_{N}\rightarrow\N(0,1)$
in distribution, as $n\rightarrow\infty$, where $\var_{p}(\cdot)$
is the variance under the sampling design for Sample A.

\end{assumption} 

We derive the asymptotic theory for $\hat{\mu}_{g,\nni}$ in the following
theorem and defer its proof to the Supplementary Material. 

\begin{theorem}\label{Thm:1}Under Assumptions \ref{asmp:MAR}\textendash \ref{asmp:regularity}
and $NN_{B}^{-1}=O(1)$, $\hat{\mu}_{g,\nni}$ has the same distribution
as $\hat{\mu}_{g,\HT}$ as $N_{B}\rightarrow\infty$. Furthermore,
under Assumption \ref{asmp:sampling}, $\hat{\mu}_{g,\nni}$ is consistent
for $\mu_{g}$, and
\begin{equation}
n^{1/2}(\hat{\mu}_{g,\nni}-\mu_{g})\rightarrow\N(0,V_{\nni}),\label{eq:asymp var-1}
\end{equation}
where
\[
V_{\nni}=\lim_{n\rightarrow\infty}\frac{n}{N^{2}}E\left[\var_{p}\left\{ \sum_{i\in A}\pi_{i}^{-1}g(Y_{i})\right\} \right].
\]

\end{theorem}Theorem \ref{Thm:1} implies that the standard point
estimator can be applied to the imputed data $\{(X_{i},Y_{i(1)}):i\in A\}$
as if the $Y_{i(1)}$'s were observed values. Let $\pi_{ij}$ be the
joint inclusion probability for units $i$ and $j$. We show in the
Supplementary Material that the direct variable estimator based on
the imputed data 
\[
\hat{V}_{\nni}=\frac{n}{N^{2}}\sum_{i\in A}\sum_{j\in A}\frac{\pi_{ij}-\pi_{i}\pi_{j}}{\pi_{i}\pi_{j}}\frac{g(Y_{i(1)})}{\pi_{i}}\frac{g(Y_{j(1)})}{\pi_{j}}.
\]
is consistent for $V_{\nni}$. 

\section{Other techniques for mass imputation\label{sec:tools}}

\subsection{$k$-nearest neighbor imputation}

Instead of using a single imputed value, we now consider fractional
imputation with $k$ imputed values for each missing outcome. Fractional
imputation is designed to reduce the variance of the final estimator
due to imputation \citep{kalton1984some,kim2004fractional}. 

Assume no matching ties, let $\mathcal{J}_{k}(i)$ be the set of $k$
nearest neighbors for unit $i$ 
\[
\mathcal{J}_{k}(i)=\left\{ l\in B:\sum_{j\in B}1_{\{d(X_{j},X_{i})\leq d(X_{l},X_{i})\}}\leq k\right\} =\left\{ i(1),\ldots,i(k)\right\} .
\]

The $k$ nearest neighbor approach to mass imputation can be described
in the following steps: 
\begin{description}
\item [{Step$\ $1.}] For each unit $i\in A$, find the $k$ nearest neighbors
from Sample B, $\mathcal{J}_{k}(i)$. Impute the $Y$ value for unit
$i$ by $\hat{\mu}_{g}(X_{i})=k^{-1}\sum_{j=1}^{k}g(Y_{i(j)})$. 
\item [{Step$\ $2.}] The $k$ nearest neighbor imputation estimator of
$\mu_{g}$ is
\[
\hat{\mu}_{g,\knn}=\frac{1}{N}\sum_{i\in A}\pi_{i}^{-1}\hat{\mu}_{g}(X_{i}).
\]
\end{description}
In the nonparametric estimation literature, researchers have investigated
the asymptotic properties of the $k$ nearest neighbor imputation
estimators extensively. See, e.g., \citet{mack1979multivariate} and
\citet{mack1981local} for early references. \citet{cheng1994nonparametric}
established root-$n$ consistency of the $k$ nearest neighbor imputation
estimator of the outcome mean when the outcome is subject to missingness.
We derive the asymptotic theory for $\hat{\mu}_{g,\knn}$ in the context
of mass imputation combining a probability sample and a big data sample
in the following theorem and defer its proof to the Supplementary
Material. 

\begin{theorem}\label{Thm:knn}Under Assumptions \ref{asmp:MAR}\textendash \ref{asmp:sampling},
$n\left(k/N\right)^{4/p}\rightarrow0$, $k/n\rightarrow0$, and $k^{2}/n\rightarrow\infty$,
\begin{equation}
n^{1/2}(\hat{\mu}_{g,\knn}-\mu_{g})\rightarrow\N(0,V_{\knn}),\label{eq:asymp var-2}
\end{equation}
where 
\[
V_{\knn}=\lim_{n\rightarrow\infty}\frac{n}{N^{2}}\left(E\left[\var_{p}\left\{ \sum_{i\in A}\pi_{i}^{-1}\mu_{g}(X_{i})\right\} \right]+E\left\{ \frac{1-\pi_{B}(X)}{\pi_{B}(X)}\sigma_{g}^{2}(X)\right\} \right),
\]
and $\pi_{B}(X)=\pr(\delta_{B}=1\mid X)$. 

\end{theorem}

If $\pi_{B}(X)$ goes to $1$, $V_{\knn}$ reduces to $\lim_{n\rightarrow\infty}\left(n/N^{2}\right)E\left[\var_{p}\left\{ \sum_{i\in A}\pi_{i}^{-1}\mu_{g}(X_{i})\right\} \right]$.
In this case, $V_{\knn}$ is smaller than $V_{\nni}$, suggesting
that $\hat{\mu}_{g,\knn}$ gains efficiency over $\hat{\mu}_{g,\nni}$.
In finite samples, \citet{beretta2016nearest} conduct a simulation
study to compare nearest neighbor imputation and $k$ nearest neighbor
imputation in the setting with independent and identically distributed
data. They found that $k$ nearest neighbor imputation with a small
$k$ outperforms nearest neighbor imputation in terms of mean squared
error.

\subsection{Generalized additive models}

Nearest neighbor imputation methods are nonparametric. On the other
hand, parametric models especially linear models are sensitive to
model misspecification. We now consider semiparametric methods for
mass imputation. Among semiparametric methods, generalized additive
models (\citealp{hastie1990GAM}) are flexible regarding model specification
of the dependence of $Y$ on $X$ by specifying the model only through
smooth functions rather than assuming a parametric relationship. We
apply generalized additive models to leverage the predictive power
of the big data sample to produce a predictive model for $Y$ given
$X$, so as to facilitate mass imputation for the probability sample. 

We assume that $g(Y_{i})$ given $X_{i}$ follows some exponential
family distribution, and
\begin{equation}
h^{-1}\{\mu_{g}(X_{i})\}=f_{1}(X_{i}^{1})+f_{2}(X_{i}^{2})+\cdots f_{p}(X_{i}^{p}),\label{eq:GAM}
\end{equation}
where $h(\cdot)$ is an inverse link function, and each $f_{k}(\cdot)$
is a smooth function of $X^{k}$, for $k=1,\ldots,p$. Model (\ref{eq:GAM})
allows for rather flexible specification of the dependence of $Y$
on $X$. The estimated function $f_{k}(X^{k})$ can reveal possible
nonlinearities of the relationship of $Y$ and $X^{k}$. 

There are several challenges in fitting model (\ref{eq:GAM}). First,
$f_{k}(x)$ is an infinite-dimensional parameter, estimation of which
often relies on some approximation. Second, we need to decide how
smooth the $f_{k}(x)$ should be to balance the trade-off between
model complexity and overfitting to the data at hand. 

To solve the first issue, a common way to approximate $f_{k}(x)$
using splines. Let $B_{m}(x)$ be the basis spline functions for $m=1,\ldots,M$
\citep{ruppert2009semiparametric}. We approximate $f_{k}(x)$ by
$f_{k}(x)=\sum_{m=1}^{M}\gamma_{m}^{k}B_{m}(x)$ with spline coefficients
$\gamma_{m}^{k}$. This leads to an approximation of model (\ref{eq:GAM}):
\begin{equation}
h^{-1}[\hat{E}\{g(Y_{i})\mid X_{i}\}]=\sum_{k=1}^{p}\sum_{m=1}^{M}\gamma_{m}^{k}B_{m}(X_{i}^{k}).\label{eq:GAM-1}
\end{equation}

In (\ref{eq:GAM-1}), a large $M$ allows for increased model complexity
and also an increased chance of overfitting; while a small $M$ may
result in an inadequate model. This trade-off is balanced by choosing
a relative large $M$ and then penalizing the model complexity in
the estimation stage (\citealp{eilers1996flexible}). Let the vector
of spline coefficients be $\gamma_{k}^{\T}=(\gamma_{1}^{k},\ldots,\gamma_{m}^{k})$
and $\gamma^{\T}=(\gamma_{1}^{\T},\ldots,\gamma_{p}^{\T})$. The estimate
$\hat{\gamma}$ is obtained by maximizing the penalized likelihood:
\begin{equation}
-2l(\gamma)+\sum_{k=1}^{p}\lambda_{k}\gamma_{k}^{\T}S_{k}\gamma_{k}\label{eq:p-lik}
\end{equation}
where $l(\gamma)$ is the log likelihood function of $\gamma$, $S_{k}$
is a matrix with the $(m,l)$th component $\int B_{m}''(x)B_{l}''(x)\de x$,
$\gamma_{k}^{\T}S_{k}\gamma_{k}$ regularizes $f_{k}$ to be smooth
for which the degree of smoothness is controlled by $\lambda_{k}$.
Given the smoothing parameter $\lambda^{\T}=(\lambda_{1},\ldots,\lambda_{p}),$
the penalized likelihood function in (\ref{eq:p-lik}) is optimized
by a penalized version of the iteratively reweighted least squares
algorithm \citep{nelder1972generalized,mccullagh1984generalized}
to obtain $\hat{\gamma}$. Regarding the choice of $\lambda,$ we
note that $\lambda$ controls the trade-off between model complexity
and overfitting, which can be estimated separately from other model
coefficients using generalized cross-validation or estimated simultaneously
using restricted maximum likelihood estimation \citep{wood2006generalized}.
In practice, the model performance is not sensitive to the choice
of the number of basis functions, but rather estimation of the smoothing
parameter is critical to control the model complexity. 

Once fitting the model, we can create an imputed value for each element
$i$ in Sample A as
\[
\hat{\mu}_{g,\gam}(X_{i})=h\{\hat{f}_{1}(X_{i}^{1})+\hat{f}_{2}(X_{i}^{2})+\cdots\hat{f}_{p}(X_{i}^{p})\},
\]
where $\hat{f}_{k}(x)=\sum_{m=1}^{M}\hat{\gamma}_{m}^{k}B_{m}(x)$
for $k=1,\ldots,p$. The mass imputation estimator based on the generalized
additive model is
\[
\hat{\mu}_{g,\gam}=\frac{1}{N}\sum_{i\in A}\pi_{i}^{-1}\hat{\mu}_{g,\gam}(X_{i}).
\]
Because in our context, the sample size of Sample B is much larger
than that of Sample A, the estimation error in the imputation model
can be negligible compared to the sampling variability of $\hat{\mu}_{g,\gam}$. 

To close this subsection, it is worth commenting on the assumption
of additive effects of $X$ in model (\ref{eq:GAM}). This assumption
may be fairly strong one. To relax the additivity assumption, we can
extend model (\ref{eq:GAM}) to include interactions through using
the tensor product basis. For example, we can include a bivariate
interaction surface $f_{12}(X^{1},X^{2})=\sum_{m=1}^{M}\sum_{l=1}^{L}\gamma_{ml}B_{m}(X^{1})B_{l}(X^{2})$. 

\section{Regression calibration\label{subsec:Regression-calibration}}

In practice, especially for government agencies, one nearest neighbor
may be preferred because of its simplicity in implementation and data
storage. We now consider another strategy to improve the efficiency
for $\hat{\mu}_{g,\nni}$ when additionally the membership to Sample
B can be determined throughout Sample A with the indicator $\delta_{B}$.
We can obtain $\delta_{B}$ by matching or directly asking about the
membership to Sample B. The key insight is that the subsample of units
in Sample  A with $\delta_{B}=1$ constitutes a second-phase sample
from Sample B, where Sample B acts as a new population.

Let $h(\delta_{B},X,Y)$ be a multi-dimensional function of $\delta_{B}$,
$\delta_{B}X$ and $\delta_{B}Y$, e.g., $h(\delta_{B},X,Y)=(\delta_{B},1-\delta_{B},\delta_{B}X,\delta_{B}Y)^{\T}$.
For simplicity of notation, we use $h_{i}$ to denote $h(\delta_{Bi},X_{i},Y_{i})$.
We can calculate the population quantity $H=N^{-1}\sum_{i=1}^{N}h_{i}$
from Sample B. This insight enables the typical calibration weighting
in survey sampling with known marginal totals. In Sample A, we treat
the imputed values as observed values, and the design weighted estimator
of $H$ is $\hat{H}_{A}=N^{-1}\sum_{i\in A}\pi_{i}^{-1}h_{i}.$ In
general, $\hat{H}_{A}$ is not equal to $H$. We can use the known
information $H$ to improve the efficiency of $\hat{\mu}_{g,\nni}$. 

This suggests the following calibration strategy. We modify the original
design weights $\{d_{i}:i\in A\}$ in $\hat{\mu}_{g,\nni}$ to a new
set of weights $\{\omega_{i}:i\in A\}$ by minimizing a distance function
\begin{equation}
\sum_{i\in A}G(\omega_{i},d_{i})=\sum_{i\in A}d_{i}\left(\frac{\omega_{i}}{d_{i}}-1\right)^{2},\label{eq:objective function}
\end{equation}
subject to the calibration constraints $N^{-1}\sum_{i\in A}\omega_{i}h_{i}=H.$
The resulting weights $\{\omega_{i}:i\in A\}$ can be called generalized
regression weights. 

The proposed estimator utilizing the new set of weights is 
\begin{equation}
\hat{\mu}_{g,\rc}=\frac{1}{N}\sum_{i\in A}\omega_{i}g(Y_{i(1)}),\label{eq:proposed estimator}
\end{equation}
which is asymptotically equivalent to a generalized regression estimator
\citep{park2012generalized}.

We derive the asymptotic theory for $\hat{\mu}_{g,\rc}$ in the following
theorem and defer its proof to the Supplementary Material. 

\begin{theorem}\label{Thm:2}Under Assumptions \ref{asmp:MAR}\textendash \ref{asmp:sampling},
\begin{equation}
n^{1/2}(\hat{\mu}_{g,\rc}-\mu_{g})\rightarrow\N(0,V_{\rc}),\label{eq:asymp var}
\end{equation}
in distribution, as $n\rightarrow\infty$, where
\[
V_{\rc}=\lim_{n\rightarrow\infty}\frac{n}{N^{2}}E\left(\var_{p}\left[\sum_{i\in A}\pi_{i}^{-1}\left\{ g(Y_{i})-h_{i}^{\T}\beta_{N}\right\} \right]\right),
\]
and $\beta_{N}=\left(\sum_{i=1}^{N}h_{i}h_{i}^{\T}\right)^{-1}\sum_{i=1}^{N}h_{i}g(Y_{i})$. 

\end{theorem} 

The calibrated estimator $\hat{\mu}_{g,\rc}$ improves the efficiency
of $\hat{\mu}_{g,\nni}$ in the sense that $V_{\rc}$ is at most as
large as $V_{\nni}$ given in Theorem \ref{Thm:1}. Moreover, $\hat{\mu}_{g,\rc}$
is robust in the sense that we do not require any modeling assumption. 

\begin{remark}[Choice of distance functions] 

Different distance functions in (\ref{eq:objective function}) can
be considered. If we choose $G(\omega_{i},d_{i})=-d_{i}\log(\omega_{i}/d_{i})$,
it leads to empirical likelihood estimation \citep{newey2004higher}.
If we choose the Kullback\textendash Leibler distance function $G(\omega_{i},d_{i})=\omega_{i}\log(d_{i}/\omega_{i})$,
it leads to exponential tilting estimation (\citealp{kitamura1997information};
\citealp{imbens1998information}; \citealp{schennach2007point}).
Under mild conditions, these procedures provide a set of weights that
is asymptotically equivalent to the set of regression weights \citep{deville1992calibration,breidt2017model}.

\end{remark} 

For variance estimation, by Theorem \eqref{Thm:2}, we construct a
consistent variance estimator for $\hat{\mu}_{g,\rc}$ as $\widehat{V}_{\rc}/n$,
where

\[
\widehat{V}_{\rc}=\frac{n}{N^{2}}\sum_{i\in A}\sum_{j\in A}\frac{\pi_{ij}-\pi_{i}\pi_{j}}{\pi_{ij}}\frac{\hat{e}_{i}}{\pi_{i}}\frac{\hat{e}_{j}}{\pi_{j}},
\]
with $\hat{e}_{i}=g(Y_{i(1)})-h_{i}^{\T}\hat{\beta}$, and 
\[
\hat{\beta}=\left(\sum_{i=1}^{N}h_{i}h_{i}^{\T}\right)^{-1}\left(\begin{array}{c}
\sum_{i=1}^{N}\delta_{Bi}g(Y_{i})\\
\sum_{i\in A}\pi_{i}^{-1}(1-\delta_{Bi})g(Y_{i(1)})\\
\sum_{i=1}^{N}\delta_{Bi}X_{i}g(Y_{i})\\
\sum_{i=1}^{N}\delta_{Bi}Y_{i}g(Y_{i})
\end{array}\right).
\]

\section{Empirical experiments\label{sec:A-simulation-study}}

In this section, we evaluate the finite sample performance of the
proposed estimator using simulation studies, one based on artificial
data and the other based on a synthetic population file from a single
month sample of the U.S. Census Bureau\textquoteright s Monthly Retail
Trade Survey.

\subsection{A simulation study\label{subsec:A-simulation-study}}

We generate the data according to the following mechanism. We first
generate a finite population $\mathcal{F}_{N}=\{X_{i}=(X_{1i},X_{2i}),Y_{i}=(Y_{1i},Y_{2i}):i=1,\ldots N\}$
with size $N=1,000,000$, where $Y_{1i}$ is a continuous outcome
and $Y_{2i}$ is a binary outcome. From the finite population, we
select a big data sample $B$ where the inclusion indicator $\delta_{Bi}\sim$Ber$(p_{i})$
with $p_{i}$ the inclusion probability for unit $i$, and we obtain
a representative sample $A$ of size $n=1,000$ using simple random
sampling. The parameters of interest are the population mean $N^{-1}\sum_{i=1}^{N}Y_{i}$
and the conditional population mean of $Y_{1}$ given $Y_{2}=1$. 

For generating the finite population, we consider linear models 
\begin{eqnarray}
Y_{1i} & = & 1+X_{1i}+X_{2i}+\alpha_{i}+\epsilon_{i},\label{eq:linearOR}\\
\pr(Y_{2i}=1\mid X_{1i},X_{2i};\alpha_{i}) & = & \logit(1+X_{1i}+X_{2i}+\alpha_{i}),\nonumber 
\end{eqnarray}
and nonlinear models 
\begin{eqnarray}
Y_{i} & = & 0.5(X_{1i}-1.5)^{2}+X_{2i}^{2}+\alpha_{i}+\epsilon_{i},\label{eq:nonlinearOR}\\
\pr(Y_{2i}=1\mid X_{1i},X_{2i};\alpha_{i}) & = & \logit\left\{ 0.5(X_{1i}-1.5)^{2}+X_{2i}^{2}+\alpha_{i}\right\} ,\nonumber 
\end{eqnarray}
where $X_{1i}\sim\N(1,1)$, $X_{2i}\sim$Exp$(1)$, $\alpha_{i}\sim\N(0,1)$,
$\epsilon_{i}\sim\N(0,1)$, and $X_{1i}$, $X_{2i}$, $\alpha_{i}$
and $\epsilon_{i}$ are mutually independent. The variables $\alpha_{i}$
induce the dependence of $Y_{1i}$ and $Y_{2i}$ even adjusting for
$X_{1i}$ and $X_{2i}$. For the big-data inclusion probability, we
also consider a logistic linear model 
\begin{equation}
\logit(p_{i})=X_{2i},\label{eq:linearlogit}
\end{equation}
and a nonlinear logistic model 
\begin{equation}
\logit(p_{i})=-3+(X_{1i}-1.5)^{2}+(X_{2i}-2)^{2}.\label{eq:nonlinearlogit}
\end{equation}
We consider the following combinations: I. (\ref{eq:linearOR}) and
(\ref{eq:linearlogit}); II. (\ref{eq:linearOR}) and (\ref{eq:nonlinearlogit});
II. (\ref{eq:nonlinearOR}) and (\ref{eq:linearlogit}); and IV. (\ref{eq:nonlinearOR})
and (\ref{eq:nonlinearlogit}) for data generating mechanisms. Therefore,
the simulation setup is a $2\times2$ factorial design with two levels
in each factor.

Kim and Wang (2018) proposed the inverse propensity score weighting
estimator using the estimated probability of selection into Sample
B and the double robust estimator which further incorporates an outcome
regression model. To evaluate the robustness and efficiency, we compare
the following estimators:
\begin{enumerate}
\item $\hat{\mu}_{\HT}$, the Horvitz\textendash Thompson estimator assuming
the $Y_{i}$'s were observed in Sample A for the purpose of benchmark
comparison;
\item $\hat{\mu}_{\ipw}$, the inverse propensity score weighting estimator,
\[
\hat{\mu}_{\ipw}=\frac{1}{N}\sum_{i\in B}\frac{1}{p_{i}(\hat{\eta})}Y_{i(1)},
\]
where $p_{i}(\eta)=\pr(\delta_{Bi}=1\mid X_{i};\eta)$ is a logistic
regression model with a linear predictor $X_{2i}$ as a working model,
and $\hat{\eta}$ is an estimator of $\eta$ based on Sample A;
\item $\hat{\mu}_{\dr}$, the double robust estimator,
\[
\hat{\mu}_{\dr}=\frac{1}{N}\sum_{i\in B}\frac{1}{p_{i}(\hat{\eta})}\left(Y_{i(1)}-X_{i}^{\T}\hat{\beta}\right)+\frac{1}{n}\sum_{i\in A}X_{i}^{\T}\hat{\beta},
\]
where $\hat{\beta}$ is the estimated regression coefficients using
(\ref{eq:linearOR}) as the working outcome regression model based
on Sample B;
\item $\hat{\mu}_{\nni}$, the nearest neighbor imputation estimator; 
\item $\hat{\mu}_{\knn}$, the $k$ nearest neighbor imputation estimator
with $k=5$; 
\item $\hat{\mu}_{\gam}$, the generalized additive model imputation estimator; 
\item $\hat{\mu}_{\rc}$, the regression calibration estimator based on
$\hat{\mu}_{\nni}$ with calibration variables $H(\delta_{B},X,Y)=(\delta_{B},1-\delta_{B},\delta_{B}X,\delta_{B}Y)^{\T}$.
\end{enumerate}
All simulation results are based on $1,000$ Monte Carlo runs. Table
\ref{tab:1} summarizes the simulation results with biases, standard
errors, and coverage rates of $95\%$ confidence intervals using asymptotic
normality of the point estiamtors. The following observations can
be made from Table \ref{tab:1}. $\hat{\mu}_{\ipw}$ has large biases
when the propensity score is misspecified. $\hat{\mu}_{\dr}$ gains
robustness over $\hat{\mu}_{\ipw}$ if one of the outcome regression
model or the propensity score is correctly specified. However, if
both models are misspecified, $\hat{\mu}_{\dr}$ has a larger bias.
$\hat{\mu}_{\nni}$ has small biases across four scenarios, which
shows its robustness. Importantly, the performance of $\hat{\mu}_{\nni}$
is close to that of $\hat{\mu}_{\HT}$ in terms of standard errors
and coverage rates, which is consistent with our theory in Theorem
\ref{Thm:1}. Moreover, as predicted by our theoretical results, $\hat{\mu}_{\knn}$
improves $\hat{\mu}_{\nni}$ in terms of efficiency. Also, $\hat{\mu}_{\gam}$
shows robustness because of the flexibility of the model specification.
The regression calibration estimator $\hat{\mu}_{\rc}$ has small
biases across all scenarios and therefore shows robustness against
model specifications for sampling score and outcome. Moreover, it
has smaller standard errors than both $\hat{\mu}_{\nni}$ and $\hat{\mu}_{\knn}$.
The coverage rates are all close to the nominal level. 
\begin{table}
\begin{centering}
{\scriptsize{}{}\caption{\label{tab:1}Simulation results: bias, standard error, and coverage
rate of $95\%$ confidence intervals under four scenarios based on
$1,000$ Monte Carlo samples. OM: outcome model; PS: propensity score
model}
} 
\par\end{centering}
\centering{}\centering{}\resizebox{\columnwidth}{!}{ %
\begin{tabular}{crrrrrrrrrrrr}
 &  &  &  &  &  &  &  &  &  &  &  & \tabularnewline
\hline 
 & Bias & S.E. & C.R. & Bias & S.E. & C.R. & Bias & S.E. & C.R. & Bias & S.E. & C.R.\tabularnewline
 & $\times10^{2}$ & $\times10^{2}$ & $\times10^{2}$ & $\times10^{2}$ & $\times10^{2}$ & $\times10^{2}$ & $\times10^{2}$ & $\times10^{2}$ & $\times10^{2}$ & $\times10^{2}$ & $\times10^{2}$ & $\times10^{2}$\tabularnewline
\hline 
 & \multicolumn{3}{c}{Scenario I} & \multicolumn{3}{c}{Scenario II} & \multicolumn{3}{c}{Scenario III} & \multicolumn{3}{c}{Scenario IV}\tabularnewline
OM & \multicolumn{3}{c}{linear } & \multicolumn{3}{c}{linear } & \multicolumn{3}{c}{nonlinear } & \multicolumn{3}{c}{nonlinear }\tabularnewline
PS & \multicolumn{3}{c}{linear} & \multicolumn{3}{c}{nonlinear} & \multicolumn{3}{c}{linear } & \multicolumn{3}{c}{nonlinear }\tabularnewline
\hline 
\multicolumn{13}{c}{Population Mean of $Y_{1}$}\tabularnewline
$\hat{\mu}_{\HT}$ & 0.2 & 6.5 & 96.0 & -0.2 & 6.4 & 94.5 & 0.61 & 15.2 & 95.7 & -0.5 & 15.6 & 93.5\tabularnewline
$\hat{\mu}_{\ipw}$ & -0.1 & 1.6 & 95.5 & 25.0 & 47.0 & 97.6 & -0.1 & 4.1 & 95.5 & 465.1 & 427.0 & 76.8\tabularnewline
$\hat{\mu}_{\dr}$ & 0.1 & 4.6 & 95.7 & 0.0 & 4.5 & 96.6 & 0.7 & 14.0 & 95.6 & 266.7 & 460.2 & 37.9\tabularnewline
$\hat{\mu}_{\nni}$ & 0.2 & 6.5 & 95.1 & -0.3 & 6.4 & 94.7 & 0.7 & 15.2 & 94.6 & -0.6 & 15.6 & 93.7\tabularnewline
$\hat{\mu}_{\knn}$ & 0.2 & 4.9 & 96.1 & -0.3 & 4.9 & 95.6 & 0.5 & 14.5 & 94.6 & -0.6 & 14.9 & 93.8\tabularnewline
$\hat{\mu}_{\gam}$ & 0.1 & 4.5 & 95.7 & -0.2 & 4.5 & 96.0 & 0.5 & 14.3 & 94.9 & -0.6 & 14.8 & 93.4\tabularnewline
$\hat{\mu}_{\rc}$ & 0.0 & 3.2 & 95.5 & -0.2 & 4.1 & 95.3 & -0.1 & 4.8 & 95.0 & 0.1 & 6.7 & 95.5\tabularnewline
\hline 
\multicolumn{13}{c}{Population Mean of $Y_{2}$}\tabularnewline
$\hat{\mu}_{\HT}$ & -0.0 & 1.5 & 96.2 & -0.0 & 1.6 & 95.1 & -0.1 & 1.6 & 95.2 & 0.1 & 1.6 & 94.4\tabularnewline
$\hat{\mu}_{\ipw}$ & 0.0 & 0.2 & 95.5 & -12.3 & 3.7 & 0.0 & 0.0 & 0.4 & 95.9 & 2.8 & 4.0 & 70.6\tabularnewline
$\hat{\mu}_{\dr}$ & -0.0 & 0.8 & 95.9 & -0.8 & 3.8 & 69.8 & -0.0 & 0.7 & 95.4 & 9.7 & 7.1 & 10.0\tabularnewline
$\hat{\mu}_{\nni}$ & 0.0 & 1.4 & 95.3 & -0.0 & 1.6 & 95.3 & -0.1 & 1.6 & 94.6 & 0.1 & 1.6 & 95.3\tabularnewline
$\hat{\mu}_{\knn}$ & 0.0 & 1.0 & 95.8 & -0.0 & 1.1 & 95.8 & -0.0 & 1.0 & 95.2 & 0.0 & 0.9 & 96.1\tabularnewline
$\hat{\mu}_{\gam}$ & -0.0 & 0.9 & 95.3 & -0.0 & 0.9 & 94.8 & -0.0 & 0.8 & 96.2 & 0.0 & 0.8 & 94.5\tabularnewline
$\hat{\mu}_{\rc}$ & 0.0 & 1.2 & 95.5 & -0.1 & 1.4 & 94.2 & -0.0 & 1.4 & 94.1 & 0.1 & 1.5 & 95.6\tabularnewline
\hline 
\multicolumn{13}{c}{Conditional Mean of $Y_{1}$ given $Y_{2}=1$}\tabularnewline
$\hat{\mu}_{\HT}$ & 0.0 & 7.3 & 95.1 & -0.3 & 7.2 & 95.2 & 0.2 & 9.3 & 95.3 & -0.1 & 9.8 & 94.1\tabularnewline
$\hat{\mu}_{\ipw}$ & -0.1 & 1.5 & 95.5 & -8.4 & 12.2 & 70.4 & -0.1 & 1.3 & 95.7 & 20.6 & 2.5 & 0.0\tabularnewline
$\hat{\mu}_{\dr}$ & 0.1 & 4.7 & 94.8 & 2.3 & 4.5 & 93.3 & 0.8 & 5.5 & 94.5 & 24.2 & 5.5 & 0.8\tabularnewline
$\hat{\mu}_{\nni}$ & -0.0 & 7.3 & 95.0 & -0.3 & 7.3 & 95.3 & 0.1 & 9.2 & 95.4 & -2.2 & 9.5 & 95.2\tabularnewline
$\hat{\mu}_{\knn}$ & -0.1 & 4.7 & 96.8 & -0.3 & 4.6 & 96.5 & 0.1 & 6.0 & 94.8 & 0.0 & 6.4 & 93.6\tabularnewline
$\hat{\mu}_{\gam}$ & 0.0 & 4.8 & 94.2 & -0.3 & 4.5 & 96.0 & -0.1 & 6.5 & 95.5 & -0.6 & 6.8 & 94.8\tabularnewline
$\hat{\mu}_{\rc}$ & -0.0 & 3.9 & 94.8 & -0.2 & 5.0 & 96.0 & -0.2 & 5.4 & 95.1 & -0.1 & 5.4 & 96.7\tabularnewline
\hline 
\end{tabular}}
\end{table}

\subsection{Monthly retail trade survey\label{sec:Real-data-application}}

To demonstrate the practical relevance, we consider the U.S. Census
Bureau\textquoteright s 2014 Monthly Retail Trade Survey (\citealp{mulry2014detecting}).
The Monthly Retail Trade Survey is an economic indicator survey whose
monthly estimates are inputs to the Gross Domestic Product estimates.
This survey selects a sample of about $12,000$ retail businesses
each month with paid employees to collect data on sales and inventories.
It employs an one-stage stratified sample with stratification based
on major industry, further substratified by the estimated annual sales
referred to as the size variable. 

For simulation purpose, according to the 2014 Monthly Retail Trade
Survey, we generate a finite population of $N=812,765$ retail businesses
with $16$ strata with a stratum identifier $h$, sales $Y$, inventories
$X$, and a size variable $Z$ on the log scale. Table \ref{tab:The-sample-allocation}
reports some summary statistics extracted from the actual survey.
We generate the inventory and size data from $X_{hi}\sim N(\mu_{X,h},\sigma_{X,h}^{2})$
and $Z_{hi}\sim N(\mu_{X,h},\sigma_{X,h}^{2})$, for $i=1,\ldots,N_{h}$
and $h=1,\ldots,16$, and the sales data from a linear model
\begin{eqnarray}
Y_{hi} & = & \beta_{0}+X_{hi}+Z_{hi}+\epsilon_{hi},\label{eq:linearOR-1}
\end{eqnarray}
and a nonlinear model
\begin{eqnarray}
Y_{hi} & =\beta_{0}+ & X_{hi}^{2}+Z_{hi}^{2}+\epsilon_{hi},\label{eq:nonlinearOR-1}
\end{eqnarray}
where $\epsilon_{hi}\sim\N(0,0.52)$. In (\ref{eq:linearOR-1}) and
(\ref{eq:nonlinearOR-1}), we specify different values for $\beta_{0}$
so that the parameter of interest, $\mu=N^{-1}\sum_{h=1}^{16}\sum_{i=1}^{N_{h}}Y_{hi}$,
matches with the true population mean $12.73$. 

\begin{table}
\caption{\label{tab:The-sample-allocation}The stratum size, sample allocation,
mean and standard error of the inventory data on the log scale extracted
from the 2014 Monthly Retail Trade Survey }
\centering{}%
\begin{tabular}{ccccccccc}
 &  &  &  &  &  &  &  & \tabularnewline
\hline 
Stratum $h$ & 1 & 2 & 3 & 4 & 5 & 6 & 7 & 8\tabularnewline
\hline 
 $N_{h}$ & 366  & 20 & 2,015  & 4,646 & 7,402 & 700 & 12,837 & 17,080\tabularnewline
 $n_{h}$ & 37 & 5 & 34 & 57 & 74 & 7 & 103 & 115\tabularnewline
$\mu_{X,h}$ & 16.8 & 16.7 & 16.6 & 16.4 & 16.1 & 15.6 & 16.0 & 15.7\tabularnewline
$\sigma_{X,h}$ & 1.1 & 0.8 & 0.4 & 0.3 & 0.4 & 0.6 & 0.4 & 0.4\tabularnewline
\hline 
Stratum $h$ & 9 & 10 & 11 & 12 & 13 & 14 & 15 & 16\tabularnewline
\hline 
$N_{h}$ & 29,808 & 2,400 & 41,343 & 57,518 & 83,465 & 95,244 & 115,028 & 342,893\tabularnewline
$n_{h}$ & 116 & 12 & 184 & 196 & 218 & 200 & 220 & 336\tabularnewline
$\mu_{X,h}$ & 15.6 & 15.5 & 15.4 & 15.1 & 14.8 & 14.5 & 13.9 & 11.5\tabularnewline
$\sigma_{X,h}$ & 0.4 & 0.3 & 0.4 & 0.4 & 0.3 & 0.7 & 0.5 & 1.1\tabularnewline
\hline 
\end{tabular}
\end{table}

We also generate a big data sample $\mathcal{S}_{B}$ where the inclusion
indicator $\delta_{hi}\sim$Ber$(p_{hi})$ with the inclusion probability
$p_{hi}$ for unit $i$ in stratum $h$. The big data sample in practice
is often available from E-commercial companies who monitor inventories
and sales for retail businesses. For the big data inclusion probability,
we consider a logistic linear model 
\begin{equation}
\logit(p_{hi})=\alpha_{0}+Z_{hi},\label{eq:linearlogit-1}
\end{equation}
and a nonlinear logistic model 
\begin{equation}
\logit(p_{hi})=\alpha_{0}+X_{hi}+Z_{hi}^{2},\label{eq:nonlinearlogit-1}
\end{equation}
where we specify different values for $\alpha_{0}$ so that the mean
inclusion probability is about $30\%$. Lastly, we generate a representative
sample $\mathcal{S}_{A}$ by stratified sampling with simple random
sampling within strata without replacement; see Table \ref{tab:The-sample-allocation}
for the sample allocation. 

We consider the seven estimators in $\mathsection$ \ref{subsec:A-simulation-study}
adopted for stratified sampling. In each mass imputed dataset, we
apply the following point estimator and variance estimator: $\hat{\mu}=N^{-1}\sum_{h=1}^{H}N_{h}\bar{y}_{n_{h}}$
with $\bar{y}_{n_{h}}$ is the sample mean of $y$ in the $h$th stratum,
$\hat{V}(\hat{\mu})=N^{-2}\sum_{h=1}^{H}N_{h}^{2}(1-n_{h}/N_{h})s_{n_{h}}^{2}/n_{h}$
with $s_{n_{h}}^{2}=(n_{h}-1)^{-1}\sum_{i=1}^{n_{h}}(y_{hi}-\bar{y}_{n_{h}})^{2}$.

Table \ref{tab:1-2} summarizes the simulation results. A similar
discussion to $\mathsection$ \ref{subsec:A-simulation-study} applies.
$\hat{\mu}_{\ipw}$ is sensitive to misspecification of the selection
model; while $\hat{\mu}_{\dr}$ has double robustness feature, which
still relies on at least one model to be correctly specified. Mass
imputation based on nearest neighbor imputation, $k$ nearest neighbor
imputation and generalized additive model shows good performances
by leveraging the representativeness of the survey sample and the
predictive power of the big data sample. In addition, if the big data
membership is known throughout the survey data, the regression calibration
estimator gains efficiency while maintaining the robustness against
model misspecification. 

\begin{table}
\begin{centering}
{\scriptsize{}{}\caption{\label{tab:1-2}Simulation results: bias, standard error, and coverage
rate of $95\%$ confidence intervals under four scenarios based on
$1,000$ Monte Carlo runs for the 2014 Monthly Retail Trade Survey.
OM: outcome model; PS: propensity score model}
} 
\par\end{centering}
\centering{}\centering{}\resizebox{\columnwidth}{!}{ %
\begin{tabular}{ccccccccccccc}
 &  &  &  &  &  &  &  &  &  &  &  & \tabularnewline
\hline 
 & Bias & S.E. & C.R. & Bias & S.E. & C.R. & Bias & S.E. & C.R. & Bias & S.E. & C.R.\tabularnewline
 & $\times10^{2}$ & $\times10^{2}$ & $\times10^{2}$ & $\times10^{2}$ & $\times10^{2}$ & $\times10^{2}$ & $\times10^{2}$ & $\times10^{2}$ & $\times10^{2}$ & $\times10^{2}$ & $\times10^{2}$ & $\times10^{2}$\tabularnewline
\hline 
 & \multicolumn{3}{c}{Scenario I} & \multicolumn{3}{c}{Scenario II} & \multicolumn{3}{c}{Scenario III} & \multicolumn{3}{c}{Scenario IV}\tabularnewline
OM & \multicolumn{3}{c}{linear } & \multicolumn{3}{c}{linear } & \multicolumn{3}{c}{nonlinear } & \multicolumn{3}{c}{nonlinear }\tabularnewline
PS & \multicolumn{3}{c}{linear} & \multicolumn{3}{c}{nonlinear} & \multicolumn{3}{c}{linear } & \multicolumn{3}{c}{nonlinear }\tabularnewline
\hline 
$\hat{\mu}_{\HT}$ & 0.0 & 4.2 & 96.1 & 0.0 & 4.2 & 96.1 & 3.3 & 85.4 & 96.2 & 3.3 & 85.4 & 96.2\tabularnewline
$\hat{\mu}_{\ipw}$ & 0.1 & 12.8 & 96.4 & -25.7 & 7.7 & 10.3 & 4.1 & 340.3 & 96.3 & -850 & 202.1 & 1.8\tabularnewline
$\hat{\mu}_{\dr}$ & 0.1 & 3.6 & 96.2 & 0.1 & 3.6 & 96.4 & 4.2 & 92.5 & 96.7 & -220 & 93.4 & 38.6\tabularnewline
$\hat{\mu}_{\nni}$ & 0.0 & 4.2  & 96.0 & 0.1 & 4.1 & 96.9 & 2.6 & 85.4 & 96.2 & 2.1 & 85.3 & 96.1\tabularnewline
$\hat{\mu}_{\knn}$ & 0.1 & 3.7 & 96.6 & 0.1 & 3.7 & 96.1 & 1.9 & 85.3 & 96.0 & 1.2 & 85.2 & 96.0\tabularnewline
$\hat{\mu}_{\gam}$ & 0.1 & 3.6 & 96.3 & 0.1 & 3.60 & 96.6 & -2.7 & 85.5 & 96.1 & -19.0 & 85.8 & 95.7\tabularnewline
$\hat{\mu}_{\rc}$ & 0.0 & 3.7 & 95.8 & 0.1 & 3.89 & 96.6 & 3.2 & 76.0 & 96.0 & 1.0 & 83.6 & 96.3\tabularnewline
\hline 
\end{tabular}}
\end{table}

\section{Discussion\label{sec:Discussion}}

Mass imputation is an important technique for survey data integration.
When the training dataset for imputation is obtained from a probability
sample, the theory of \citet{kim2012combining} can be directly applied.
If the training dataset is a non-probability sample and its size is
huge, we have shown in this article that various nonparametric methods
can be used for mass imputation, and the estimation error in the imputation
model can be safely ignored, under the assumption that the sampling
mechanism for training data is missing at random in the sense of \citet{rubin1976inference}.
If the sampling mechanism is believed to be not missing not at random,
imputation techniques can be developed under the strong model assumptions
for the sampling mechanism (e.g. \citealp{riddles2016propensity,morikawa2018note}).
Also, when the training dataset has a hierarchical structure, multi-level
models can be used to develop mass imputation. This is closely related
to unit-level small area estimation in survey sampling \citep{rao2015small}.
These are topics for future research. 

\section*{Supplementary Material}

Supplementary material includes the proofs for three theorems.

\bibliographystyle{dcu}
\bibliography{C:/Dropbox/bib/ci,C:/Dropbox/bib/pfi_MIsurvey_v6}

\pagebreak{}

\global\long\def\theequation{S\arabic{equation}}
 \setcounter{equation}{0}

\global\long\def\thelemma{S\arabic{lemma}}
 \setcounter{lemma}{0}

\global\long\def\theexample{S\arabic{example}}
 \setcounter{equation}{0}

\global\long\def\thesection{S\arabic{section}}
 \setcounter{section}{0}

\global\long\def\thetheorem{S\arabic{theorem}}
 \setcounter{equation}{0}

\global\long\def\thecondition{S\arabic{condition}}
 \setcounter{equation}{0}

\global\long\def\theremark{S\arabic{remark}}
 \setcounter{equation}{0}

\global\long\def\thestep{S\arabic{step}}
 \setcounter{equation}{0}

\global\long\def\theassumption{S\arabic{assumption}}
 \setcounter{assumption}{0}

\global\long\def\theproof{S\arabic{proof}}
 \setcounter{equation}{0}

\global\long\def\theproposition{S{proposition}}
 \setcounter{equation}{0} 
\begin{center}
\textbf{\huge{}{}{}Supplementary material }{\huge{}{}}\\
 {\huge{}{} }\textbf{\huge{}{} \bigskip{}
 }\textbf{\large{}{}{}by Shu Yang and Jae Kwang Kim}{\large{}{}
} 
\par\end{center}

\bigskip{}
 \bigskip{}

\setcounter{page}{1}

\section{Proofs}

\subsection{Proof for Theorem \ref{Thm:1}}

For a given $X_{i}=x$ in Sample A, we show that $X_{i(1)}$ converges
to $x$ in probability as $N_{B}\rightarrow\infty.$ To show this,
consider for any $\epsilon>0$,
\begin{eqnarray}
\pr\{d(X_{i(1)},x)>\epsilon\} & = & \pr\{d(X_{j},x)>\epsilon\ \forall j\in B\}\nonumber \\
 & = & \left[\pr\{d(X_{j},x)>\epsilon\}\right]^{N_{B}}.\label{eq:S0}
\end{eqnarray}
By Assumption \ref{asmp:overlap}, $x$ is in the support of $X$
in Sample B. This leads to $\pr\{d(X_{j},x)<\epsilon\}>0$ and $\pr\{d(X_{j},x)>\epsilon\}<1$.
Therefore, (\ref{eq:S0}) converges to zero, and $X_{i(1)}$ converges
to $x$ in probability as $N_{B}\rightarrow\infty.$ 

Given $X_{i}=x$, for any continuous and bounded $g(y)$,
\begin{eqnarray*}
E\{g(Y_{i(1)})\mid X_{i}=x\} & = & E[E\{g(Y_{i(1)})\mid X_{i(1)}\}\mid X_{i}=x]\\
 & = & E\{\mu_{g}(X_{i(1)})\mid X_{i}=x\}\\
 & \rightarrow & E\{\mu_{g}(X_{i})\mid X_{i}=x\}\\
 & = & E\left\{ g(Y_{i})\mid X_{i}=x\right\} ,
\end{eqnarray*}
in probability as $N_{B}\rightarrow\infty$, where $\rightarrow$
follows from the fact that $\mu_{g}(x)$ is bounded and continuous.
Then, by Portmanteau Lemma \citep{klenke2006probability}, $Y_{i(1)}\rightarrow Y_{i}\mid(X_{i}=x)$
in distribution as $N_{B}\rightarrow\infty$. By Assumption \ref{asmp:MAR},
$g(Y_{i(1)})\mid X_{i}\rightarrow\mu_{g}(X_{i})+e_{g}^{*}(X_{i})$
in distribution as $N_{B}\rightarrow\infty$, where $e_{g}^{*}(X_{i})$
has the same distribution as $\left\{ g(Y_{i})\mid X_{i}\right\} -\mu_{g}(X_{i})$. 

We now show that for $i\neq j\in A$, $e_{g}^{*}(X_{i})$ and $e_{g}^{*}(X_{j})$
are conditionally independent, given data $\mathcal{O}_{A}$. It is
sufficient to show that $\pr\left\{ i(1)=j(1)\right\} \rightarrow0$
as $N_{B}\rightarrow\infty$; in other words, the same unit can not
be matched for unit $i$ and unit $j$ with probability $1$. This
can be shown using (\ref{eq:S0}) with $\epsilon=\min_{i\neq j\in A}||X_{i}-X_{j}||$. 

Therefore, conditional on data $\mathcal{O}_{A}$,
\[
\hat{\mu}_{g,\nni}=\frac{1}{N}\sum_{i\in A}\pi_{i}^{-1}g(Y_{i(1)})\rightarrow\frac{1}{N}\sum_{i\in A}\pi_{i}^{-1}g(Y_{i})=\hat{\mu}_{g,\HT}
\]
in distribution as $N_{B}\rightarrow\infty$. This completes the proof
for Theorem \ref{Thm:1}.

Let
\begin{equation}
\tilde{V}_{\nni}=\frac{n}{N^{2}}\sum_{i\in A}\sum_{j\in A}\frac{\pi_{ij}-\pi_{i}\pi_{j}}{\pi_{i}\pi_{j}}\frac{g(Y_{i})}{\pi_{i}}\frac{g(Y_{j})}{\pi_{j}}.\label{eq:V1}
\end{equation}
Then, $\tilde{V}_{\nni}$ is consistent for $V_{\nni}$. 

Similar to the above argument, for $i,j\in A$, conditional on data
$\mathcal{O}_{A}$, $g(Y_{i(1)})g(Y_{j(1)})\rightarrow g(Y_{i})g(Y_{j})$
as $N_{B}\rightarrow\infty$. Therefore, conditional on data $\mathcal{O}_{A}$,
\begin{equation}
\hat{V}_{\nni}=\frac{n}{N^{2}}\sum_{i\in A}\sum_{j\in A}\frac{\pi_{ij}-\pi_{i}\pi_{j}}{\pi_{i}\pi_{j}}\frac{g(Y_{i(1)})}{\pi_{i}}\frac{g(Y_{j(1)})}{\pi_{j}}\rightarrow\tilde{V}_{\nni},\label{eq:V2}
\end{equation}
in distribution as $N_{B}\rightarrow\infty$. Combining (\ref{eq:V1})
and (\ref{eq:V2}), $\hat{V}_{\nni}$ is consistent for $V_{\nni}$. 

\subsection{Proof for Theorem \ref{Thm:knn}}

To investigate the asymptotic properties of $\hat{\mu}_{g,\knn}$,
we re-express
\[
\hat{\mu}_{g}(x)=\frac{\sum_{j\in B}K_{R_{x}}(x-X_{j})g(Y_{j})}{\sum_{j\in B}K_{R_{x}}(x-X_{j})},
\]
where
\[
K_{h}(u)=\frac{1}{h^{p}}K\left(\frac{u}{h}\right),\ \ K(u)=0.5I(||u||\leq1),
\]
and the bandwidth $h=R_{x}$ is the random distance between $x$ and
its furthest among the $k$ nearest neighbors. Therefore, $\hat{\mu}_{g,\knn}$
can be viewed as a kernel estimator incorporating a data-driven bandwidth. 

In the literature, asymptotic properties of the $k$ nearest neighbor
imputation estimator have been studied extensively. The result shown
in the following lemma on $k$ nearest neighbor imputation is extracted
from \citet{mack1981local}.

\begin{lemma}\label{lemma:knn-1}Under Assumptions \ref{asmp:MAR}\textendash \ref{asmp:regularity},
\begin{equation}
N^{-1}\sum_{j=1}^{N}\delta_{B,j}K_{R_{x}}(x-X_{j})g(Y_{j})=f(x)\pi_{B}(x)\mu_{g}(x)+O_{p}\left\{ \left(\frac{k}{N}\right)^{2/p}+\frac{1}{k}\right\} .\label{eq:knn-1}
\end{equation}

\end{lemma}

We now express
\begin{eqnarray*}
\hat{\mu}_{g,\knn} & = & \frac{1}{N}\sum_{i=1}^{N}\pi_{i}^{-1}\delta_{A,i}\mu_{g}(X_{i})+\frac{1}{N}\sum_{i=1}^{N}\pi_{i}^{-1}\delta_{A,i}\left\{ \hat{\mu}_{g}(X_{i})-\mu_{g}(X_{i})\right\} .
\end{eqnarray*}
Let $T_{N}=N^{-1}\sum_{i=1}^{N}\pi_{i}^{-1}\delta_{A,i}\left\{ \hat{\mu}_{g}(X_{i})-\mu_{g}(X_{i})\right\} $.
To study the properties for $T_{N}$, we first look at $\hat{\mu}_{g}(x)$,
which can be expressed as
\[
\hat{\mu}_{g}(x)=\frac{h_{N}(x)}{f_{N}(x)},
\]
where $h_{N}(x)\equiv N^{-1}\sum_{j=1}^{N}\delta_{B,j}K_{R_{x}}(x-X_{j})g(Y_{j})$
and $f_{N}(x)\equiv N^{-1}\sum_{j=1}^{N}\delta_{B,j}K_{R_{x}}(x-X_{j})$.
By the result in Lemma \ref{lemma:knn-1}, we obtain
\begin{eqnarray*}
h_{N}(x) & = & f(x)\pi_{B}(x)\mu_{g}(x)+O_{p}\left\{ \left(\frac{k}{N}\right)^{2/p}+\frac{1}{k}\right\} \\
f_{N}(x) & = & f(x)\pi_{B}(x)+O_{p}\left\{ \left(\frac{k}{N}\right)^{2/p}+\frac{1}{k}\right\} .
\end{eqnarray*}
Now, by a Taylor expansion, we obtain
\begin{eqnarray*}
\hat{\mu}_{g}(x)-\mu_{g}(x) & = & \frac{h_{N}(x)}{f_{N}(x)}-\mu_{g}(x)\\
 & = & \frac{1}{f(x)\pi_{B}(x)}\left\{ h_{N}(x)-f(x)\pi_{B}(x)\mu_{g}(x)\right\} \\
 &  & -\frac{f(x)\pi_{B}(x)\mu_{g}(x)}{\{f(x)\pi_{B}(x)\}^{2}}\left\{ f_{N}(x)-f(x)\pi_{B}(x)\right\} +O_{p}\left\{ \left(\frac{k}{N}\right)^{2/p}+\frac{1}{k}\right\} \\
 & = & \frac{1}{f(x)\pi_{B}(x)}\left\{ h_{N}(x)-f_{N}(x)\mu_{g}(x)\right\} +O_{p}\left\{ \left(\frac{k}{N}\right)^{2/p}+\frac{1}{k}\right\} .
\end{eqnarray*}
Therefore, we obtain
\[
T_{N}=\frac{1}{N^{2}}\sum_{i=1}^{N}\frac{\delta_{A,i}}{\pi_{i}}\frac{1}{f(X_{i})\pi_{B}(X_{i})}\sum_{j=1}^{N}\delta_{B,j}K_{R_{X_{i}}}(X_{i}-X_{j})\{g(Y_{j})-\mu_{g}(X_{i})\}+O_{p}\left\{ \left(\frac{k}{N}\right)^{2/p}+\frac{1}{k}\right\} .
\]
Under the assumption in Theorem \ref{Thm:knn}, it is easy to drive
that $\left(k/N\right)^{2/p}+1/k=o(n^{-1/2}),$ and therefore,
\[
T_{N}=\frac{1}{N^{2}}\sum_{i=1}^{N}\frac{\delta_{A,i}}{\pi_{i}}\frac{1}{f(X_{i})\pi_{B}(X_{i})}\sum_{j=1}^{N}\delta_{B,j}K_{R_{X_{i}}}(X_{i}-X_{j})\{g(Y_{j})-\mu_{g}(X_{i})\}+o_{p}(n^{-1/2}).
\]
We then express $T_{N}$ in a form of U-statistics (\citealp{van1998asymptotic};
Ch. 12):
\[
T_{N}=\frac{1}{N(N-1)}\sum_{i=1}^{N}\sum_{j\neq i}h(Z_{i},Z_{j})+o_{p}(n^{-1/2}),
\]
where $Z_{i}=(X_{i},Y_{i},\delta_{A,i},\delta_{B,i})$ and 
\begin{eqnarray*}
h(Z_{i},Z_{j}) & = & \frac{1}{2}\left[\frac{\delta_{A,i}\delta_{B,j}}{\pi_{i}}\frac{1}{f(X_{i})\pi_{B}(X_{i})}K_{R_{X_{i}}}(X_{i}-X_{j})\{g(Y_{j})-\mu_{g}(X_{i})\}\right.\\
 &  & +\left.\frac{\delta_{A,j}\delta_{B,i}}{\pi_{j}}\frac{1}{f(X_{j})\pi_{B}(X_{j})}K_{R_{X_{j}}}(X_{j}-X_{i})\{g(Y_{i})-\mu_{g}(X_{j})\}\right]\\
 & \equiv & \frac{1}{2}(\zeta_{ij}+\zeta_{ji}).
\end{eqnarray*}
Now, by Lemma \ref{lemma:knn-1}, we obtain
\begin{eqnarray*}
E(\zeta_{ij}\mid Z_{i}) & = & E\left[\frac{\delta_{A,i}\delta_{B,j}}{\pi_{i}}\frac{1}{f(X_{i})\pi_{B}(X_{i})}K_{R_{X_{i}}}(X_{i}-X_{j})\{g(Y_{j})-\mu_{g}(X_{i})\}\mid Z_{i}\right]\\
 & = & \frac{\delta_{A,i}}{\pi_{i}}\frac{1}{f(X_{i})\pi_{B}(X_{i})}E\left[\delta_{B,j}K_{R_{X_{i}}}(X_{i}-X_{j})\{g(Y_{j})-\mu_{g}(X_{i})\}\mid Z_{i}\right]\\
 & = & O\left\{ \left(\frac{k}{N}\right)^{2/p}+\frac{1}{k}\right\} ,
\end{eqnarray*}
and
\begin{eqnarray*}
E(\zeta_{ji}\mid Z_{i}) & = & E\left[\frac{\delta_{A,j}\delta_{B,i}}{\pi_{j}}\frac{1}{f(X_{j})\pi_{B}(X_{j})}K_{R_{X_{j}}}(X_{j}-X_{i})\{g(Y_{i})-\mu_{g}(X_{j})\}\mid Z_{i}\right]\\
 & = & \delta_{B,i}E\left(E\left[\frac{\delta_{A,j}}{\pi_{j}}\frac{1}{f(X_{j})\pi_{B}(X_{j})}K_{R_{X_{j}}}(X_{j}-X_{i})\{g(Y_{i})-\mu_{g}(X_{j})\}\mid R_{X_{j}},Z_{i}\right]\mid Z_{i}\right)\\
 & = & \frac{\delta_{B,i}}{\pi_{B}(X_{i})}\{g(Y_{i})-\mu_{g}(X_{i})\}+O\left\{ \left(\frac{k}{N}\right)^{2/p}+\frac{1}{k}\right\} .
\end{eqnarray*}
Therefore, by the theory of U-statistics, we obtain
\begin{eqnarray*}
T_{N} & = & \frac{2}{N}\sum_{i=1}^{N}E\{h(Z_{i},Z_{j})\mid Z_{i}\}+o_{p}(n^{-1/2})\\
 & = & \frac{1}{N}\sum_{i=1}^{N}\frac{\delta_{B,i}}{\pi_{B}(X_{i})}\{g(Y_{i})-\mu_{g}(X_{i})\}+o_{p}(n^{-1/2}).
\end{eqnarray*}
Combining the above results leads to
\begin{eqnarray}
\hat{\mu}_{g,\knn}-\mu_{g} & = & \frac{1}{N}\sum_{i=1}^{N}\left\{ \pi_{i}^{-1}\delta_{A,i}\mu_{g}(X_{i})-\mu_{g}(X_{i})\right\} \nonumber \\
 &  & +\frac{1}{N}\sum_{i=1}^{N}\left\{ \frac{\delta_{B,i}}{\pi_{B}(X_{i})}-1\right\} \{g(Y_{i})-\mu_{g}(X_{i})\}+o_{p}(n^{-1/2}).\label{eq:knn}
\end{eqnarray}
Then, the asymptotic results in Theorem \ref{Thm:knn} follow by Assumptions
\ref{asmp:MAR}\textendash \ref{asmp:sampling} and (\ref{eq:knn}).

\subsection{Proof for Theorem \ref{Thm:2}}

The consistency and asymptotic normality of $n^{1/2}\hat{\mu}_{g,\nni}$
follow by the standard arguments under Assumptions \ref{asmp:MAR}\textendash \ref{asmp:sampling}.
The remaining is to show that the asymptotic variance of $n^{1/2}\hat{\mu}_{g,\nni}$
is $V_{\nni}.$ 

Using the distance function $G(\omega_{i},d_{i})=d_{i}(\omega_{i}/d_{i}-1)^{2}$
in (\ref{eq:objective function}), the minimum distance estimation
leads to generalized regression estimation \citep{park2012generalized}.
Therefore, we express
\begin{eqnarray}
n^{1/2}\hat{\mu}_{g} & = & \frac{n^{1/2}}{N}\sum_{i\in A}\omega_{i}g(Y_{i(1)})\nonumber \\
 & = & \frac{n^{1/2}}{N}\sum_{i\in A}\pi_{i}^{-1}g(Y_{i(1)})-\frac{n^{1/2}}{N}\left(\sum_{i\in A}\pi_{i}^{-1}h_{i}^{\T}\beta_{N}-\sum_{i=1}^{N}h_{i}^{\T}\beta_{N}\right)+o_{p}(n^{-1/2}).\label{eq:S6}
\end{eqnarray}
Similar to the argument in the proof for Theorem \ref{Thm:1}, we
express
\begin{eqnarray}
n^{1/2}\hat{\mu}_{g} & = & \frac{n^{1/2}}{N}\sum_{i\in A}\pi_{i}^{-1}g(Y_{i(1)})-\frac{n^{1/2}}{N}\left(\sum_{i\in A}\pi_{i}^{-1}h_{i}^{\T}\beta_{N}-\sum_{i=1}^{N}h_{i}^{\T}\beta_{N}\right)+o_{p}(n^{-1/2})\nonumber \\
 & = & \frac{n^{1/2}}{N}\sum_{i\in A}\pi_{i}^{-1}\left\{ g(Y_{i(1)})-h_{i}^{\T}\beta_{N}\right\} +\frac{n^{1/2}}{N}\sum_{i=1}^{N}h_{i}^{\T}\beta_{N}+o_{p}(n^{-1/2}).\label{eq:S5}
\end{eqnarray}
It is straightforward to show the variance of the second term in (\ref{eq:S5})
is negligible given $nN^{-1}=o(1)$. Following the arguments in the
proof for Theorems \ref{Thm:1} and \ref{Thm:knn}, $g(Y_{i(1)})$
has the asymptotic distribution as $g(Y_{i})$ given the data $\mathcal{O}_{A}$
from Sample A. Therefore, the asymptotic variance of $n^{1/2}\hat{\mu}_{g}$
is
\[
V_{\rc}=\lim_{n\rightarrow\infty}\var\left[\frac{n^{1/2}}{N}\sum_{i\in A}\pi_{i}^{-1}\left\{ g(Y_{i})-h_{i}^{\T}\beta_{N}\right\} \right].
\]

\end{document}